\documentclass[superscriptaddress,twocolumn,floats,showpacs,prl,amsmath,amssymb,floatfix,nofootinbib,balancelastpage]{revtex4}

\input epsf
\usepackage{psfig}% Include figure files

\newcommand{\spin}[1]{\,{}_{#1}^{\vphantom{m}}}

\begin{document}

\title{Cosmic 21-cm Delensing of Microwave Background Polarization \\ and the Minimum Detectable Energy Scale of Inflation}
\author{Kris Sigurdson}
\email{ksigurds@tapir.caltech.edu}
\affiliation{California Institute of Technology, Mail Code 130-33, Pasadena, CA
91125}
\author{Asantha Cooray}
\email{acooray@uci.edu}
\affiliation{California Institute of Technology, Mail Code 130-33, Pasadena, CA
91125}
\affiliation{Department of Physics and Astronomy, 4186 Frederick Reines Hall, University of California, Irvine, CA 92697}

\begin{abstract}
The curl (B) modes of cosmic microwave background (CMB) polarization anisotropies are a unique probe of the primordial background of inflationary gravitational waves (IGWs).
Unfortunately, the B-mode polarization anisotropies generated by gravitational waves at recombination are confused with those generated by the mixing of gradient-mode (E-mode) and B-mode polarization anisotropies as CMB photons propagate through the Universe and are gravitationally lensed.
We describe here a method for delensing CMB polarization anisotropies using observations of anisotropies in the cosmic 21-cm radiation emitted or absorbed by neutral hydrogen atoms at redshifts 10 to 200. While the detection of cosmic 21-cm anisotropies at high resolution is challenging, a combined study with a relatively low-resolution (but high-sensitivity) CMB polarization experiment could probe inflationary energy scales well below the Grand Unified Theory (GUT) scale of $10^{16}$~GeV --- constraining models with energy scales below $10^{15}$~GeV (the detectable limit derived from CMB observations alone).  The ultimate theoretical limit to the detectable inflationary energy scale via this method may be as low as $3 \times 10^{14}$~GeV.
\end{abstract}


\maketitle

\noindent \emph{Introduction--- }The curl (B) modes of cosmic microwave background (CMB) polarization anisotropies are a unique probe of the primordial background of cosmological gravitational waves \cite{KamKosSte97}. 
At these long wavelengths, inflation \cite{Gut81} is the only known mechanism to causally generate such
a background of gravitational waves. Since the amplitude of these inflationary gravitational waves (IGWs) is proportional to ${\cal V}$, the value of the inflaton potential $V(\varphi)$ during inflation, the amplitude of gravitational-wave induced B-mode polarization anisotropies directly constrains the energy scale of inflation ${\cal V}^{1/4}$ (see, for example, Ref.~\cite{KamKos99}).
While the experimental sensitivity to B-mode polarization can be improved, the expected signal is contaminated by foreground effects \cite{tucci}.  The main confusion to the detection of B-mode polarization anisotropies generated by IGWs at recombination is the mixing of gradient-mode (E-mode) and B-mode anisotropies via gravitational lensing \cite{ZalSel98}.  

In this \emph{Letter} we propose a new method for separating lensing-induced B modes from the IGW signal using observations of anisotropies in the cosmic 21-cm radiation emitted or absorbed by neutral hydrogen atoms at redshifts 10 to 200.  While the detection of cosmic 21-cm anisotropies at high resolution is challenging, a combined study with CMB polarization data could probe inflationary energy scales well below the Grand Unified Theory (GUT) scale of $10^{16}$ GeV --- constraining inflationary models with energy scales below $10^{15}$ GeV.  The ultimate theoretical limit to the minimum detectable energy scale of inflation via this method could reach as low as $3 \times 10^{14}$~GeV.

\noindent \emph{Gravitational Lensing--- } Lensing induces a remapping of the polarization field at the last-scattering surface ${}_{\pm} X({\hat{\bf n}})$ such that ${}_{\pm} \tilde X({\hat{\bf n}}) =   {}_{\pm}X[{\hat{\bf n}} + \nabla\phi({\hat{\bf n}}) ]$ is the observed polarization field, where
$\spin{\pm} X = Q\pm i U$ are linear combinations of the Stokes
parameters $Q$ and $U$ and ${\alpha}({{\hat{\bf n}}})=\nabla\phi({\hat{\bf n}}) $ is the lensing deflection angle.  Here,
\begin{eqnarray}
\phi({\hat{\bf n}};z_s)&=&- 2 \int_0^{r(z_s)} dr'
\frac{r-r'}{r' r}
\Phi (\hat{{\bf n}},r') \,
\label{eqn:lenspotential}
\end{eqnarray}
is the deflection potential, a line-of-sight projection of the gravitational potential $\Phi$ to redshift $z_s$.  The total lensing potential $\phi({\hat{\bf n}}) \equiv \phi({\hat{\bf n}}; z_{\rm CMB})$ is this quantity evaluated at $z_s \rightarrow z_{\rm CMB} \approx 1100$.

Using the flat-sky approximation and the E-mode/B-mode decomposition \cite{KamKosSte97}, the lensed B-mode polarization power spectrum, in the relevant limit $C_l^{BB} << C_l^{EE}$, is 
\begin{equation}
\tilde{C}_{l}^{BB} = C_{l}^{BB} + \int \frac{d^2 {\bf l}'}{(2 \pi)^2}
\left[ {\bf l}'' \cdot {\bf l}' \right]^2 \sin^2(2\theta_{l}')
C_{l''}^{\phi \phi} C_{l'}^{EE} \, ,
\label{eqn:CBB}
\end{equation}
where ${\bf l}'' = {\bf l} - {\bf l}'$. The second term in this expression is the lensing confusion in the B-mode map which must be
separated from $C_l^{BB}$ --- the IGW signal.  Here, $C_l^{\phi \phi}$ is the angular power spectrum of the total deflection potential and is simply related to a weighted projection of the matter power spectrum \cite{ZalSel98};  $C_l^{\phi \phi}(z_s)$ is the incomplete power spectrum out to source redshift $z_s < z_{\rm CMB}$.  

Unlike the B modes generated by tensor perturbations (the IGWs), $C_l^{EE}$ is dominated by larger amplitude scalar perturbations.  The expected few-percent conversion of E modes creates a large signal in the B-mode power spectrum \cite{ZalSel98}. For tensor-to-scalar ratios ${\cal T}/{\cal S}$ below $2.6 \times 10^{-4}$ or, since ${\cal V}^{1/4} = 3.0 \times 10^{-3} ({\cal T}/{\cal}S)^{1/4}m_{Pl}$ \cite{turner}, ${\cal V}^{1/4}$ below $4.6 \times 10^{15}$~GeV the IGW signal is completely confused by the lensing contaminant \cite{lewis}.
To bypass this limit one must separate the lensing induced B-modes from those due to IGWs. 
Clearly, the lensing confusion could be exactly removed if one knew the three-dimensional distribution of mass out to the CMB last-scattering surface. 
However, as our goal is a measurement of $C_{l}^{BB}$, knowledge of the projected quantity $\phi({\hat{\bf n}})$ is sufficient. One way to estimate $\phi({\hat{\bf n}};z_s)$ is by using quadratic estimators or maximum likelihood methods to statistically infer the deflection-angle field given some lensed random field $\tilde{\chi}({\hat{\bf n}})$ at source redshift $z_s$.  For instance, arcminute resolution CMB temperature and polarization maps could be used to make such an estimate \cite{lensing,Hu01,HuOka02}.   Another way to estimate $\phi({\hat{\bf n}};z_s)$ is by observing the weak-lensing distortions of the shapes of objects of a known average shape at source redshift $z_s$.  
We note here that observations of the weak lensing of galaxies, which have $z_{s} \sim 1$--$2$, can not be used to delense CMB maps because a large fraction of the lensing contamination (55\% at $l=1000$) comes from structure at $z > 3$.  Higher source redshifts are required for effective delensing.  We first review the potential observational signatures of the cosmic 21-cm radiation, and then discuss the methods for delensing the CMB outlined above.

\noindent \emph{Cosmic 21-cm Radiation--- }  Neutral atoms kinetically decouple from the thermal bath of CMB photons at $z \sim 200$ and cool adiabatically as $T_{\rm g} \propto (1+z)^2$ \cite{peebles}.  Since the spin temperature of the hydrogen atoms remains collisionally coupled to $T_g$ these atoms resonantly absorb CMB photons at $\lambda_{21}=21.1$~cm --- the hyperfine transition of the ground state of hydrogen.  The cosmic 21-cm radiation is thus first observable in absorption by low-frequency radio telescopes which could detect brightness-temperature fluctuations at wavelength $\lambda=\lambda_{21}(1+z)$ \cite{shaver,loeb,bharadwaj}.  During reionization, the neutral gas distribution is likely to be complex due to the first luminous sources \cite{barkana} and cosmic 21-cm signatures shift to emission \cite{zaldarriaga}.  Yet, even before reionization, it is possible the 21-cm sky is brightened by emission from neutral hydrogen gas contained in minihalos with masses $\sim 10^3$--$10^7$ M$_{\odot}$ \cite{shapiro}.   

Like the CMB, the statistics of the high-$z$ absorption fluctuations are expected to be Gaussian and quadratic estimators of the lensing potential, described below, could be straightforwardly adapted to reconstruct the deflection field. If the statistics of the 21-cm fluctuations during reionization can be understood the lensing of the 21-cm emission from that era might also provide a useful probe of the lensing potential.  The most promising (but futuristic) possibility involves using the shape statistics of high-redshift minihalos to infer the lensing potential.

\noindent \emph{Quadratic Estimators--- } Quadratic estimators can be used to extract lensing information from the gravitationally-lensed field $\tilde{\chi}({\hat{\bf n}})$ of some intrinsic field $\chi({\hat{\bf n}})$ at redshift $z_s$.  The quadratic form \mbox{${\mathbf \nabla}\cdot[\chi({\hat{\bf n}}) {\mathbf \nabla}\chi({\hat{\bf n}})]$} provides an estimate of the deflection angle at position ${{\hat{\bf n}}}$ on
the sky given the $\tilde{\chi}$ anisotropy map.  For the CMB, the quantity $\tilde{\chi}$ could be the temperature anisotropies \cite{Hu01}, the polarization anisotropies, or some combination of both \cite{HuOka02}.  
The brightness temperature fluctuations in the 21-cm transition of neutral hydrogen from redshifts 10 to 200 could similarly be used \cite{cooray3}.

In Fourier space, the quadratic estimator for the deflection potential is
\begin{align}
\hat{\phi}({\bf l};z_s) = {\cal Q}_l(z_s) \int
     \frac{d^2{\bf l}'}{(2\pi)^2} \left({\bf l}
     \cdot {\bf l}' {C}_{l'}^{\chi \chi} +{\bf l} \cdot {\bf l}''
     {C}_{l''}^{\chi \chi}\right) \frac{\chi(l')\chi(l'')}{2 T_{l'}^{\chi
     \chi} T_{l''}^{\chi \chi}} \, ,  
\end{align}
where
$C_l^{\chi \chi }$ is the unlensed power spectrum and \mbox{$T_l^{\chi \chi}= \tilde{C}_l^{\chi \chi } + N_l^{\chi \chi}$}
is to total power spectrum, including lensing corrections
and a noise power spectrum $N_l^{\chi \chi}$.
The expectation value of the deflection-potential estimator $\langle
\hat{\phi}({\bf l};z_s) \rangle$ (the ensemble average over realizations of the random field $\chi$) is just $\phi({\bf l};z_s)$. Here,
\begin{equation}
\label{E:Nl}
     \left[{\cal Q}_l(z_s)\right]^{-1} = \int \frac{d^2{\bf l}'}{(2\pi)^2} \frac{\left({\bf l} \cdot
     {\bf l}' C_{l'}^{\chi \chi} +{\bf l} \cdot {\bf l}''
     C_{l'}^{\chi \chi}\right)^2}{2 T_{l'}^{\chi \chi}
     T_{l''}^{\chi \chi}} \, .
\end{equation}
is the noise power spectrum associated with a quadratic reconstruction of $C_l^{\phi \phi}(z_s)$ using the field $\tilde{\chi}$ \cite{Hu01}.

\noindent \emph{Partial Delensing Bias--- } An estimate of $\phi({\hat{\bf n}})$ can be used to delense the CMB B-mode polarization map.  In the limit $z_s \rightarrow z_{\rm CMB}$ (conventional CMB delensing) the extraction of $C_l^{BB}$ from the delensed map is limited by the noise introduced during delensing.  The residual contamination of the B-modes is given by the second term of Eq.~(\ref{eqn:CBB}) with the replacement $C_l^{\phi \phi} \rightarrow {\cal Q}_{l}$. However if $z_s < z_{\rm CMB}$ this noise is not necessarily the factor limiting a measurement of the IGW signal.  Using $\hat{\phi}({\hat{\bf n}};z_s)$ as a proxy for $\phi({\hat{\bf n}})$ to delense the map leaves a residual lensing contamination not due to noise.  Accounting for this partial delensing bias ${\cal B}_l(z_s) \equiv C_l^{\phi \phi} - C_l^{\phi \phi}(z_s)$ (due to the difference in source redshift between the lensed field $\tilde{\chi}({\hat{\bf n}})$ and the CMB) the residual contamination of the B-mode power spectrum is instead the second term of Eq.~(\ref{eqn:CBB}) with $C_l^{\phi \phi} \rightarrow {\cal B}_l(z_s)+{\cal N}_{l}(z_s)$.  This is true whether $\hat{\phi}({\hat{\bf n}};z_s)$ is estimated using quadratic estimators or by some other method.  Here, ${\cal N}_{l}(z_s)$ is the residual noise power spectrum of the deflection potential due to noise associated with the delensing process --- for quadratic reconstruction ${\cal N}_{l}(z_s) = {\cal Q}_l(z_s)$. If the deflection potential is reconstructed from a line, as is the case for cosmic 21-cm radiation, the source redshift is exactly know and ${\cal B}_l(z_s)$ can be reliably estimated.

\noindent \emph{Quadratic~Reconstruction--- } Unlike the CMB anisotropies, which lack power on angular scales below a few arcminutes due to Silk damping, the cosmic 21-cm anisotropies extend to much higher values of $l$ (limited by the Jeans wavelength of the gas) and peak in amplitude at higher values of $l$ \cite{loeb,bharadwaj}.  Additionally, measurements of cosmic 21-cm anisotropies in different frequency bins provide several estimates of essentially the same deflection field.

\begin{figure}[t]
\centerline{\psfig{file=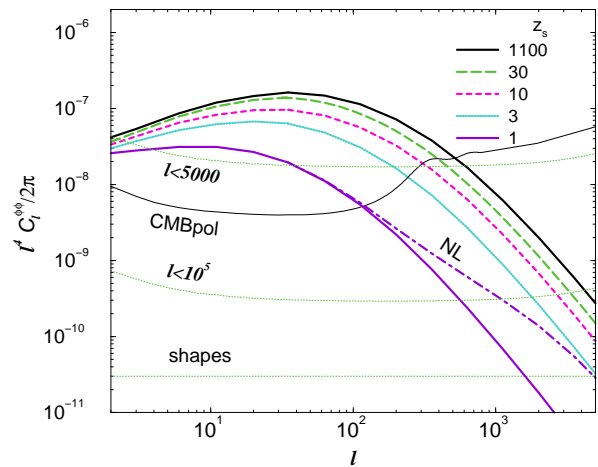,width=3.026in,angle=-90}}
\caption{Shown is angular power spectrum of the deflection potential as a function of source redshift $z_{s}$.
The curves labeled `$l < 5000$' and `$l <10^5$' are the estimated noise levels for quadratic reconstruction using 21-cm anisotropies in 40 \mbox{0.5~MHz} bins centered around $z_{s} \approx 30$.  We assume a noise power spectrum with $T_{sys}=3000 K$ at 46 MHz, that $l_{\rm max} f_{\rm cov} \approx 15$, and a year of integration \cite{zaldarriaga}.
The curve labeled shapes shows the residual noise curve in a scenario where shear is directly measured using resolved minihalos in a 1 MHz bandwidth about $z_{s} \approx 30$. Also shown is the noise levels for
a CMB reconstruction of deflections with the planned CMBpol mission assuming
a 3 arcminute beam, a noise level of 1$\mu$K $\sqrt{\rm sec}$, and a year of integration.
}
\label{deflec}
\end{figure}

As shown in Fig.~\ref{deflec}, we estimate a 21-cm experiment centered around $z_s \sim 30$ with a 20 MHz coverage in frequency space capable of observing anisotropies out to $l \sim 5000$ would have an ${\cal N}_l(z_s)$ higher than the planned CMBpol mission.  In this case residual confusion arises from noise rather than bias.  A quadratic reconstruction of the deflection field using this type of measurement, in conjunction with a CMB polarization experiment, could detect ${\cal T}/{\cal S} \gtrsim 2.5 \times 10^{-5}$ or ${\cal V}^{1/4} > 2.6 \times 10^{15}$~GeV.  A next-generation 21-cm experiment capable of observing anisotropies out to $l \sim 10^5$ with the same central redshift and bandwidth would have a ${\cal N}_l(z_s)$ an order of magnitude below CMBpol and now be limited largely by the bias ${\cal B}_l(z_s)$.  Paired with CMB polarization observations, this type of measurement could detect ${\cal T}/{\cal S} \gtrsim 1.0 \times 10^{-6}$ or ${\cal V}^{1/4} > 1.1 \times 10^{15}$~GeV --- comparable to very high sensitivity and resolution future CMB observations alone \cite{hirata}.  However, as the CMB data would not need to be used to reconstruct the deflection field a much lower resolution CMB experiment would suffice.  Furthermore, if lensing information need not be extracted from the CMB observations, the optimal observing strategy is to integrate over a few square degree patch of the sky as proposed in Ref.~\cite{Kamionkowski:1997av}.  Such a CMB experiment could thus be ground based.

\noindent \emph{Other Methods--- }  If minihalos bright in 21-cm emission exist in the early Universe, just as galaxy
shapes are sheared by weak gravitational lensing so will be the shapes of these minihalos. 
Ellipticity information obtained from such 21-cm minihalos could be used to reconstruct the projected potential out to high $z$ \cite{pen2}.  Based on the dark-matter halo mass function, we expect roughly a surface density of $10^{11}$/sr of such minihalos at $z \sim 30$ for a bandwidth of 1 MHz with masses between 10$^5$ and 10$^7$ M$_{\odot}$.  A typical halo of mass $10^6$ M$_{\odot}$ has a characteristic projected angular size of $\sim 60$ milliarcseconds.  If resolved, then techniques currently applied to measure shear in background galaxies in the low-$z$ Universe could be adapted for this application.  Regardless of the exact method, a deflection potential reconstructed from high-redshift 21-cm observations could then be used to delense CMB B-mode maps.

\noindent \emph{Bias-Limited Delensing--- }  
To understand to what extent bias-limited reconstructions, where the residual lensing contamination is dominated by ${\cal B}_l(z_s)$, would result in the removal of lensing confusion we have calculated the residual B-mode power spectrum after correcting for the modified lensing kernel when $z_{s} < z_{\rm CMB}$.  We have adapted the formalism of Ref.~\cite{knox} to estimate the smallest detectable background of IGWs and the resulting limits are summarized in Fig.~2.  While knowing the projected mass distribution  out to $z_{s}=1$ does not allow the confusion to be reduced significantly, if it is known to $z_{s}=10$ the confusion is reduced by an order of magnitude and the minimum detectable energy scale of inflation is reduced below the limit derived using quadratic CMB statistics \cite{knox}. A lensing-source redshift $z_{s} \gtrsim 30$ would be required to improve beyond the practical $1.1 \times 10^{15}$~GeV limit of the more sophisticated maximum likelihood method with high resolution and sensitivity CMB polarization observations \cite{hirata}.  However we emphasize here that in the case where a bias-limited reconstruction of the deflection field exists for $z_{s} \sim 30$, a very high-resolution CMB polarization experiment is not necessary, and a similarly sensitive experiment with a lower resolution could do the same job. Thus, it is conceivable that the Inflationary Probe of NASA's Beyond Einstein Program can be designed in combination with a cosmic 21-cm radiation experiment.  Such a cosmic 21-cm radiation experiment could be extremely exciting in its own right \cite{loeb,Profumo:2004qt}. 
\begin{figure}[t]
\centerline{\psfig{file=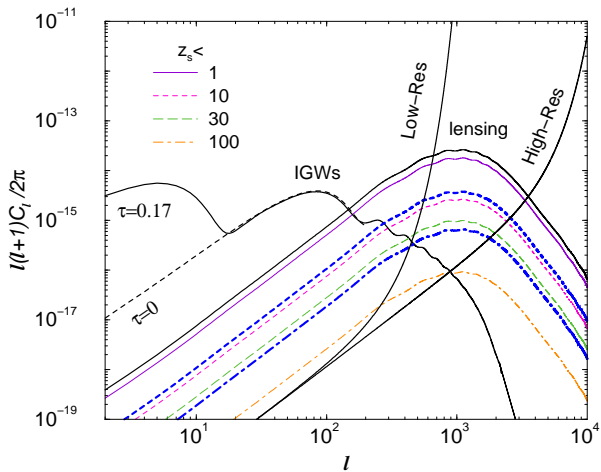,width=3.115in,angle=-90}}
\caption{Shown are power spectra of CMB B-mode polarization. The curve labeled `IGWs' 
is the IGW contribution  to B-modes assuming
a tensor-to-scalar ratio of 0.1 with (solid line; $\tau=0.17$) and without (dashed line) reionization.
The curve labeled `lensing' is the total lensing confusion to B-modes. Thin lines show the
residual B-mode lensing contamination for bias-limited delensing out to $z_{s}$.
Thick lines show previous estimates of the residual confusion from CMB experiments alone using quadratic estimators with an ideal noise-free experiment (dashed line), and likelihood methods using a high resolution/sensitivity experiment (dot-dashed). The noise curve of latter this experiment, with 2 arcminute beams and a pixel noise of 0.25 $\mu$K-arcminute, is the curve labeled `High-Res'.  Bias-limited lensing information to $z_{s} \gtrsim 10$
improves upon the limit to the IGW amplitude based on quadratic statistics, and the likelihood level can be reached with $z_{s} \sim 30$.  If $z_{s} \sim 100$ an additional order-of-magnitude in the IGW amplitude could be probed.  The curve labeled `Low-Res' is the noise curve for a lower resolution CMB polarization experiment with 30 arcminute beams sufficiently sensitive to detect IGWs when paired with a cosmic 21-cm lensing reconstruction.  For very efficient delensing other foregrounds, such as patchy reionization \cite{Santos:2003jb}, might dominate confusion.
}
\label{spectra}
\end{figure}
For a bias-limited reconstruction out to a $z_s \sim 100$, the limit on the tensor-to-scalar ratio is $7.0 \times  10^{-8}$ or  ${\cal V}^{1/4} > 6.0 \times 10^{14}$ GeV.  For $z_s \sim 200$, the maximum redshift where 21-cm fluctuations are expected to be nonzero, one could probe down to ${\cal V}^{1/4} > 3 \times 10^{14}$ GeV.  

\noindent \emph{Discussion--- }  While obtaining bias-limited measurements out to $z_s \approx 100$ is a daunting task, with many experimental and theoretical obstacles to overcome, there is great interest in detecting the fluctuations in the cosmic 21-cm radiation at $z \sim 10 - 200$ for their own sake.
The observational study of 21-cm fluctuations, especially during and prior to the era of reionization, is now being pursued by a variety of low frequency radio interferometers such as the Primeval Structure Telescope 
(PAST \cite{pen}), the Mileura Widefield Array (MWA), and the Low Frequency Array (LOFAR \cite{rott}). Planned interferometers such as the Square Kilometer Array (SKA) will improve both sensitivity and low-frequency coverage. 

While certain models of inflation, those related to Grand Unified Theories (GUTs), are expected to have an energy scale ${\cal V}^{1/4}$ between \mbox{$10^{15}$~GeV} and \mbox{$10^{16}$~GeV} there are certainly other possibilities.  For instance, some supersymmetric theories of inflation have energy scales of several times $10^{14}$~GeV \cite{RossAdams}.  New methods, such as the idea of using observations of the cosmic 21-cm radiation to delense the CMB B-mode polarization suggested here, are needed to push the minimum detectable energy scale of inflation below $10^{15}$~GeV and discriminate between between physical theories at the highest energy scales.

\smallskip
\noindent \emph{Acknowledgments--- } KS acknowledges the support of a
Canadian NSERC Postgraduate Scholarship. AC acknowledges the support of the Sherman Fairchild foundation.  This work was
supported in part by NASA NAG5-9821 and DoE DE-FG03-92-ER40701.

\end{document}